\documentclass[
reprint,
nofootinbib,
 amsmath,amssymb,
pra,
]{revtex4-2}

\usepackage{titlesec} 

\usepackage{graphicx}
\usepackage{dcolumn}
\usepackage{bm}
\usepackage{hyperref}
\usepackage[mathlines]{lineno}

\usepackage{braket}
\usepackage{nicefrac}

\begin{document}

\preprint{APS/123-QED}

\title{Hybrid Trapping of $^{87}$Rb Atoms and Yb$^{+}$ Ions in a Chip-Based Experimental Setup}

\author{Abasalt Bahrami}
 \email{abasalt@ucla.edu}
 \altaffiliation[Also at ]{Department of Electrical and Computer Engineering, University of California, Los Angeles}
\author{Matthias M\"uller}%
\author{Ferdinand Schmidt-Kaler}%

\affiliation{%
Institut f\"ur Physik, Universit\"at Mainz, Staudingerweg 7, 55128 Mainz, Germany}

\date{\today}

\begin{abstract}

Hybrid quantum systems that unite laser-cooled trapped ions and ultracold quantum gases in a single experimental setup have opened a rapidly advancing field of study, including Quantum chemistry, polaron physics, quantum information processing and quantum simulations. We present a fully developed and tested ion trap chip and propose a flat chip trap that can be placed beneath the ion trap. This design substantially addresses the difficulties specific to hybrid traps and features well-aligned chips that allow for independent adjustment of the depth of the atomic trap and the confinement and positioning of ions. The ion trap has been successfully tested with linear ion crystals of Yb$^{+}$  and neutral $^{87}$Rb were also loaded into a mMOT a few millimeters under the ion trapping region.
\end{abstract}

\maketitle


\section{\label{sec:Introduction}Introduction}

 To advance in the field of hybrid quantum systems, we employ a combination of trapping methods for atomic ions and neutral atoms. The trapped ions provide highly controllable quantum systems, making them a valuable platform for a wide range of applications, including quantum information~\cite{Haffner.2008, Singer.2010}, high-resolution spectroscopy, and tests of fundamental physics~\cite{Safronova:2018}. On the other hand, interactions between neutral atoms primarily result from short-range van der Waals forces, which range from one to several angstroms.

 Atomic ions are frequently confined using conventional quadrupole ion traps, also known as linear Paul traps~\cite{Paul:1958,Raizen:1992}. These traps utilize radio frequency fields to create a trapping potential for ions and allow for precise control of the ion motion. In contrast, neutral atoms can be confined using various techniques such as magneto-optical traps (MOTs)\cite{Migdall:1985}, dipole traps utilizing magnetic fields\cite{Pritchard.1983}, or far-detuned laser light~\cite{Kuppens.2000}. By combining these different trapping methods, we aim to create a hybrid system that can exploit the unique properties of both trapped ions and neutral atoms for various applications.

The temperature of a confined atomic cloud is typically in the nanokelvin (nK) range, while the temperature of ions confined in a Paul trap is typically in the millikelvin (mK) range. This presents a possibility of achieving sub-millikelvin temperatures for an ion crystal by means of thermalization with the atomic cloud~\cite{Krych:2013,Zipkes:2010b, Meir:2016}.

Precise positional control of both atomic and ionic constituents is a crucial aspect of experiments utilizing hybrid atom-ion systems. These versatile many-body quantum systems possess a wide range of potential applications, including quantum simulations~\cite{Monroe:2012, Bloch:2012, Grimm:2000, Bloch:2008a} and optical frequency standards~\cite{Ludlow:2015}. A significant technical challenge in experimental studies involving mixtures of ultracold atoms and ions is the integration of trapping technologies for both species into a single apparatus, enabling spatial overlap of atoms and ions. These hybrid systems provide novel platforms for investigating quantum many-body physics~\cite{Doerk:2010,Bissbort:2013,Schurer:2016}, atom-ion interactions in cold regimes~\cite{Grier:2009,Schmid:2010}, cold chemistry~\cite{Hall:2011,Rellergert:2011,Ratschbacher:2012}, and offer new opportunities for applications~\cite{Harter:2013,Cot'e:2000a,Atom-Ion-Rev:2014,Atom-Ion-Rev:2015,Cot'e:2016,Zhang:2017,Atom-Ion-Rev:2017,Trimby.2022}.

	The precision of quantum gates is limited by the presence of electric noise near the surface of the ion trap. This can be mitigated by cooling the ions via collisions with an atomic bath~\cite{Brownnutt:2015}. An ion crystal submerged in an ultracold cloud of fermionic atoms may also serve as a quantum simulator of crystalline solids~\cite{Feynman:1982}, in which the trapped ions form a periodic lattice and induce band structures in the atomic ensemble, with the atoms acting as electrons.

In hybrid systems, the atomic properties interact with the vibrations of the ionic crystal, creating a simulation of a solid-state system with improved performance on trapped ions. To study the interaction between atoms and ions, it is important to reach the so-called quantum or $s$-wave regime~\cite{Idziaszek:2011}. However, a significant challenge in achieving this is the limitations of ion trapping potentials which restrict the achievable low temperatures (below mK) in hybrid atom-ion systems. Specifically, the micromotion of ions trapped in a radio-frequency (RF) trap can lead to heating during short-range (Langevin) collisions with atoms. Research has shown that the lowest temperatures can be reached for the largest ion-atom mass ratios $m_{i}/m_{a}$\cite{Cetina:2012}. For example, by controlling the DC electric field and a mass ratio of $m_{i}/m_{a}\approx 29$, it may be possible to enter the $s$-wave regime in a Yb$^+$/Li hybrid system. So far, experimental studies have been limited to certain combinations of atoms and ions, such as Rb/Ba$^+$, Rb/Rb$^+$, Rb/Yb$^+$, Rb/Sr$^+$, and Li/Ca$^+$, for ultracold atom clouds. Additionally, other combinations of atoms and ions, such as Rb/Ca$^+$, Yb/Yb$^+$, Ca/Ba$^+$, Ca/Yb$^+$, Na/Na$^+$, Rb/K$^+$, Cs/Rb$^+$, and Na/Ca$^+$, have been studied with atoms cooled in a MOT. In our work, the combination of Rb/Yb$^+$\cite{Abasalt:2019} can be studied in both MOT and ultracold regimes in a single setup, therefore we use for the ions a surface Paul trap\cite{chiaverini.2005} and for the atoms a mMOT configuration~\cite{Reichel:1999}.
	
In this study, we outline the experimental design for hybrid experiments and examine the trapping of atoms in proximity to the location of the trapped ion crystal. To accomplish this, Rubidium atoms are loaded into a mirror magneto-optical trap (mMOT) and the fluorescence of the cold atoms is captured by a CCD camera.

\section{\label{sec:UHV}UHV system integration and characterization}
The experimental setup utilized in this study employs an ultra-high vacuum (UHV) system that has been evacuated to extremely low pressures of $2\times10^{-10}$,mbar using a combination of an ion-getter element and a non-evaporative pump (NEG) (NEXTorr\textsuperscript{\textregistered}) D200-5 NEG - ION combination pump 200l/s H2). The chip-trap, which is used to trap the ions, is mounted upside down on a CF63 flange that provides several electrical feedthroughs (Hositrad: 1x p/n 16802-01-W Sub-D Feedthrough, 2x p/n 9216-08-) for connecting the trap to other equipment. The ion trap surface is located in the precise center of the vacuum chamber, providing ideal optical access for imaging and manipulation (Kimball Physics: MCF800M-SphSq-G2E4C4 - 4$\times$8CF, 4$\times$4.5CF, 4$\times$2.75CF). The Yb oven, which is used to heat the ions to the appropriate temperature, is connected to the CF40 flange which has a high current input capability (1x p/n 9216-08-W).

Both flanges support equipment carriers, which include atom dispensers (AMD SAES: 5G0125 - RB/NF/3.4/12 FT10+10), that are used to introduce atoms into the trap. One dispenser is located directly behind the trap and serves as the primary source of atoms, while the second dispenser is placed on the ion source carrier and is only used as a reserve in case insufficient atoms are trapped. Since Rb is highly flammable, the atom dispensers are sealed and must be activated by heating them to a specific temperature. To mitigate the risk of contaminating the trap surface with atoms from the source, the primary source of Rb atoms is installed on the backside of the chip trap.

To observe and analyze the atoms and ions, high numerical aperture (NA) objectives and inverse view ports are used to bring the imaging equipment closer to the trapping region, thus increasing the resolution of the images captured. In-house constructed magnetic field coils, which are used to generate the quadrupole magnetic field, are mounted on the CF40 view ports at a $45^\circ$ angle relative to the surface of the trap chip. This allows for precise manipulation and control of the trapped ions and atoms.

\section{\label{sec:Iontrap}Design and fabrication of the chip-based hybrid atom-ion trap}
\begin{figure*}
\centering
\includegraphics[width=.8\linewidth]{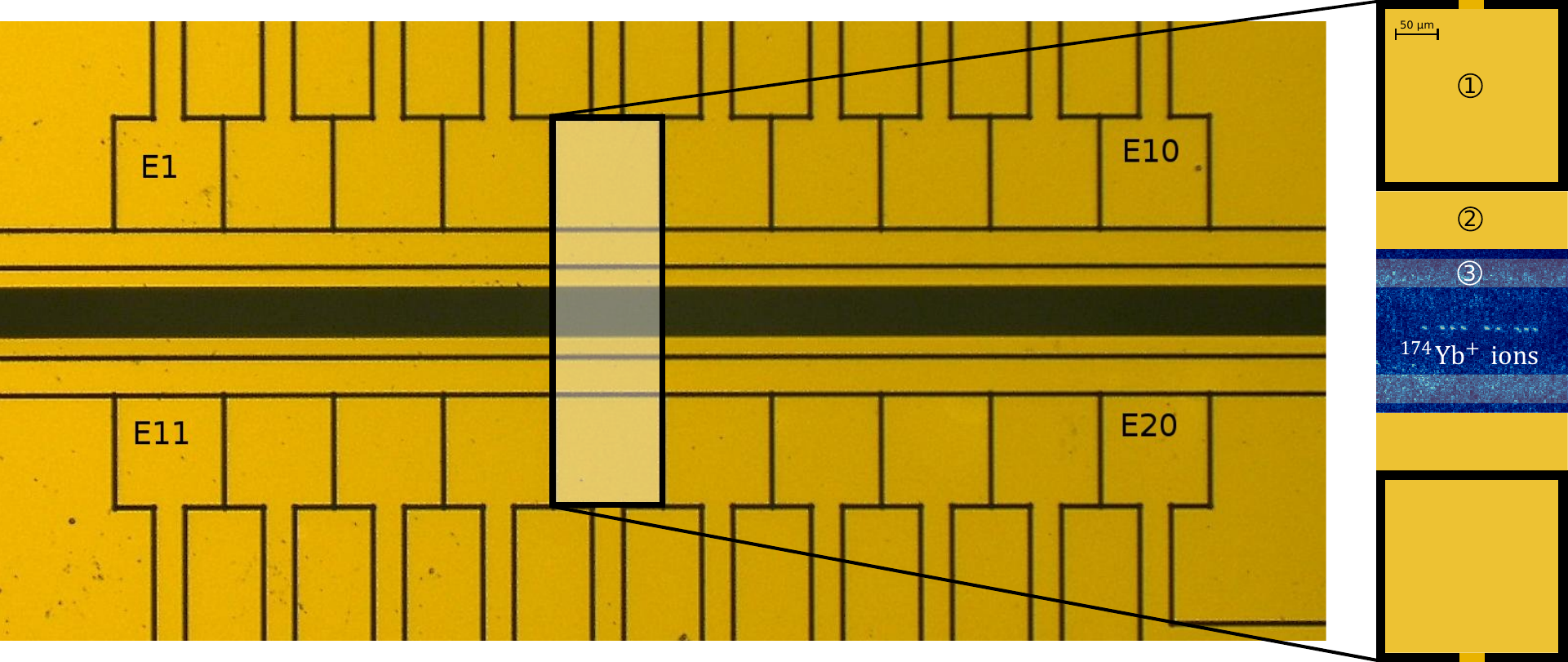}
\caption{The image above is a magnified optical microscope image of the microfabricated surface trap used in our experimental setup. The chip has dimensions of 9$\times$4.5\,mm$^{2}$ and a thickness of 500\,$\mu$m. It features 21 static voltage electrodes, including 20 with a size of 200$\times$200\,$\mu$m$^{2}$, which are used for radial confinement (E01 - E20), one long, symmetric F-shaped rail for RF confinement and one inner compensation electrode, which extends axially and symmetrically along a slit of 100\,$\mu$m width and 5\,mm length (E21). This slit is employed to load Rb atoms from the dispenser positioned directly behind the ion trap. The isolation between the electrodes is approximately 10\,$\mu$m wide and 50\,$\mu$m deep, which is large enough to prevent electrical breakdown at 100-200\,V$_{pp}$. In the experiments described here, Yb$^{+}$ ion crystals are trapped and confined along the trap axis (z-direction). The false color CCD image, captured with an exposure time of 1.3,s, depicts nine $^{174}$Yb$^{+}$ ions in a linear crystal that is trapped with corresponding trap frequencies of ($\omega_{x},\omega_{a})=2\pi\times(406, 110)$\,kHz. Dark ions observed in the image are $^{172}$Yb$^{+}$. Each pixel in the image corresponds to 1.09(7)\,$\mu$m, providing a highly detailed and precise view of the ion crystal. This microfabricated surface trap, with its intricate design and advanced manufacturing process, is essential in achieving the high-precision and high-stability trapping and manipulation of ions required for the experiments.}
\label{fig:iontrap}
\end{figure*}

	The ion trap chip utilized in this experiment was procured from the Quantum Information with ions group at Berkeley University\footnote{http://www.physics.berkeley.edu/research/haeffner/}. It is a state-of-the-art segmented planar ion trap with microstructured electrodes, and has a trapping height of 100\,$\mu$m. The chip is 9\,mm in length, 4.5\,mm in width and 500\,$\mu$m in thickness. It is equipped with a loading slit, which is 100\,$\mu$m wide and 6.5,mm long, located in the middle of the chip and is used to introduce atoms into the trap (Fig. \ref{fig:iontrap}).

The chip is fabricated using a complex process that involves etching the structure of the electrodes onto a fused silica substrate using a combination of laser attenuation and HF-etching (hydrofluoric acid, which has a strong corrosive effect on SiO$_{2}$). The etched structure is then covered with four layers of metal, specifically titanium (20\,nm), gold (150\,nm), titanium (20\,nm) and gold (150\,nm), which are applied in sequence. This advanced manufacturing process results in a high-precision and high-performance ion trap chip that is capable of trapping and manipulating ions with great accuracy and stability.

In our experimental setup, ions are confined within the node line of the quadrupole field created by the RF electrode. Additionally, ions or ion crystals are confined along the trap z-axis by a DC harmonic oscillator potential~\cite{Leibfried:2003,Bermudez:2017}. The RF and DC fields are independently adjusted to control the position and alignment of the ion crystals, the inter-ion distances, and the trapping frequencies in all directions. The challenge is to find the appropriate control voltages that match the experimental protocol and ensure that the local minimum of the DC potential aligns with the RF node line, in order to achieve a position where the ion micromotion is compensated.

The total effective potential used for trapping the ions is the sum of a time-independent potential generated by the trap DC electrodes and a sinusoidal varying part, known as the pseudopotential, that is driven by an RF voltage source. The position of the ion along the trap axis can be precisely controlled by adjusting the DC electric fields \cite{Abasalt:2019,Kaushal.2020}. This allows for the manipulation and control of the ions in the trap with high precision and stability, essential for the success of the experiments. The ion trap is firmly affixed to the atomic trap using UV glue (EPO-TEK\textsuperscript{\textregistered} OG142-112). The glue is applied with precision to the edges of the ion trap while ensuring that it does not seep between the two traps. The adhesive is hardened using UV laser light, which is applied three times for 10\,s. Electrical conductors between the chips are connected using a wire bonder with specific bonding parameters: power $P=350$\,mW, time $t=200$\,ms, force $F=60$\,cN and $200$\,$\mu$m of gold wire. To provide enough power in the RF electrode, a helical resonator is integrated, which has a frequency of $\nicefrac{\Omega}{2\pi}=11.12$\,MHz and provides a peak-to-peak voltage of about 110\,V.

\subsection{\label{sec:YbLaser}Laser alignment for the trapped ion species Yb$^{+}$}

Trapping, cooling and imaging of Yb ions requires three lasers.The ionization of Yb is done via a two level scheme. A laser at 398.9\,nm excites the neutral Yb atom to the $^1$P$_1$ state. The second step is done with a laser at 369\,nm, which ionizes the Yb atom. The same laser beam is used to doppler cool and image the ion. This laser drives the $^2$S$_{1/2}\leftrightarrow ^2$P$_{1/2}$ dipole transition and is blue detuned. In $0.5\%$ of the decays the ion falls into meta stable $^2$D$_{3/2}$ state. To bring the ion back to the cooling cycle a third laser - the repump laser - at 935\,nm is used. It brings the ion to the short living $^3$D[3/2]$_{1/2}$ which decays back to the ground state.

The blue beams from the diode lasers are overlapped and coupled to a UV polarization-maintaining (PM) fiber. The repump laser is coupled to an infrared (IR) PM fiber. Afterwards, all beams are combined via a mirror that reflects the UV light and transmits the IR light (Thorlabs M254C45: $\varnothing$1\,inch UVFS Cold Mirror, AOI: 45$^\circ$). Behind the mirror, a 200\,mm acromat lens is placed.

The focal point of the beams is located below the ion trap and the combined beam is aligned about 100\,$\mu$m below the ion chip. The focus sizes of the UV beams are 30\,$\mu$m and the repump beam is 100,$\mu$m. Some UV light is reflected by the ion trap and can be observed with the EMCCD camera (Andor Luca). This light assists in visualizing the microstructures of the planar ion chip and in focusing the camera on the surface of the trap. The reflected light is relatively faint in comparison to the fluorescence signal of the ions, thus it does not interfere with the measurements of signals from the ions. A more detailed description of the ion trap operations can be found in our paper \cite{Abasalt:2019}.

\subsection{\label{OpticalCav} Stable reference cavity and frequency locking}
Our external optical resonator is comprised of a flat and concave mirror (Altechna, partially reflective concave mirror with a radius of curvature ROC$=250$\,mm) arranged in a hemispherical configuration. The mirrors are partially reflective coated and provide a reflectivity of $R=99.0(2)$\,\% for the specific wavelengths. The plano-concave cavity mirror is mounted on an assembly of two custom-made ring-shaped piezoelectric elements (Ferroperm, Pz26, $P_\text{max}=10$\,\nicefrac{W}{cm$^2$}). The thermal expansion of both elements cancel each other due to their arrangement. A custom-made Zerodur-block (Schott AG Advanced Optics, ZERODUR, expansion class: 0) with a borehole serves as the mount, resulting in a cavity length of $L$=100\,mm that is insensitive to slight temperature fluctuations of the system, thanks to its small coefficient of thermal expansion ($\alpha(0^\circ-50^\circ)=0\pm 0.020\times10^{-6}$\,K$^{-1}$). To further stabilize and decouple the cavity from the external environment, it is placed in a vacuum chamber with $P<1.33\times 10^{-8}$\,mbar. The entrance window (Thorlabs GmbH, WG11050, AR coated for 650-1050,nm and 250-700\,nm, respectively) is inclined at an angle of $5^\circ$ with respect to the beam path to prevent reflections superimposing with the cavity modes. A small fraction of the laser light emitted by the ECDL is coupled into the cavity. A specialized CCD camera (Logitech C525 HD webcam USB) is utilized to monitor the transmitted signal of the cavity modes, while a photodiode (Thorlabs GmbH, PDA10A-EC - Si Fixed Gain Detector) provides monitoring of the back reflection from the cavity \cite{Joger.2018}. This reference cavity can also be used for Pound Drever Hall stabilization of the laser frequencies. Typical drift rates are less than $4.8$\,MHz, ensuring reliable trapping of ions.

\section{\label{sec:IntChip}Characterization of the Hybrid Trap}
\begin{figure}
		\centering
		\includegraphics[width=1\linewidth]{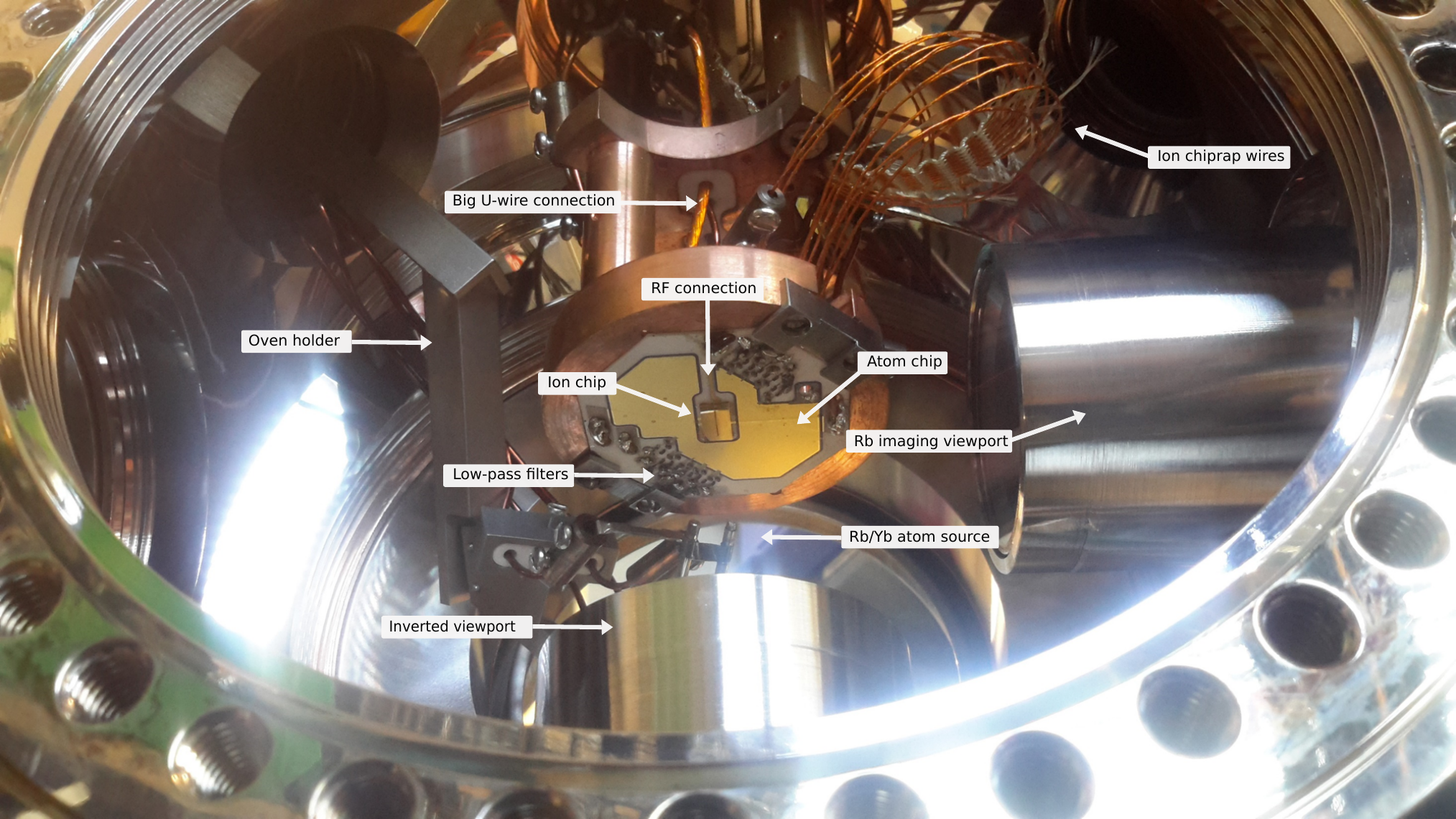}
		\caption{A visual representation of the central part of our experimental setup when installed within an ultra-high vacuum chamber. This image showcases various vital elements of the setup such as ovens, the atom-ion chip and wire bonds.}
		\label{fig:CompleteTrap}
		\end{figure}

The experimental setup comprises a linear segmented ion trap (as detailed in Section \ref{sec:Iontrap}) and an array of trapping devices for neutral Rb atoms, including a mirror-magneto-optical trap (mMOT) for loading and cooling, supplementary current-carrying wires for atom transport, and a magnetic trap for enhanced confinement (as depicted in Fig. \ref{fig:CompleteTrap}). Positioned beneath the atom chip is a large, U-shaped wire necessary for the formation of a secondary magneto-optical trap. The wires on the atom chip exhibit two distinct geometries, namely U-shaped and Z-shaped. The atom chip itself is fabricated utilizing thick film technology, enabling the printing of ultra-high vacuum compatible, sub-millimeter scale electrical circuits on an alumina substrate (Al$_{2}$O$_{3}$). The filter board, developed at the Faculty of Physics at the University of Siegen, is incorporated into the design. The utilization of thick film technology allows for the implementation of multiple circuit layers separated by insulating layers. Additionally, the chip incorporates a filter board for the direct current control electrodes of the ion trap, which incorporates 3.38\,MHz low-pass filters (with capacitance of 4.7\,nF and resistance of 10\,$\Omega$) for all direct current electrodes to mitigate RF pickups in proximity to the trap drive. Electrical bonding wires connect the DC and RF voltage from the octagonal structure to the ion trap chip. The overall configuration of the hybrid trap is an octagon with multiple conductive layers.

The overall dimensions of the chip are 45\,mm in outer diameter and 1\,mm in total height. The Z-shaped wire has a height of 0.08\,mm and width of 0.6\,mm. The magnetic field generated by the wire can be approximated as that of an ideal finite wire. The depth of the trap is determined by the bias field, its gradient, and the curvature of the magnetic field of the wire, and is calculated to be 2.3\,mK. As the trapping area is limited by the surface of the ion trap at a position of 0.6\,mm, the trap depth is estimated to be 273\,$\mu$K. As such, temperature is not a limiting factor in this system. The final trap frequencies are calculated to be approximately $\nicefrac{\omega}{2\pi} \sim (1.17, 1.17, 0.084)$\,KHz. 

A primary objective of this experiment is to investigate the interactions of atoms with ions through induced dipole moments. To this end, it is crucial to establish a characteristic range of the atom-ion interaction, which is defined by the length scale $R^{}=\sqrt{2C_{4}/\hbar^{2}}$. For these interactions, the ionic wave packet length $l_{a}=\sqrt{\hbar/m_{a}\omega_{a}}$ must be commensurate with $R^{}$. The specific mixture of $^{171}$Yb$^{+}$ and $^{87}$Rb atoms utilized in this experiment results in $R^{*}=306$\,nm, necessitating a trap frequency of 1235\,Hz. This value is closely matched by the trapping frequencies that can be achieved by applying a current of 15 A. The corresponding single atom wave packet sizes are $l_{a}$=(315.25, 315.67, 1174,76)\,nm. In contrast to traditional hybrid trap designs, this particular trap design offers improved trapping stability and a streamlined infrastructure, making it an ideal foundation for further advancements.

In order to properly affix the two chips together, the ion trap is adhered to the center of the atom trap utilizing a UV-curable adhesive (specifically, Epoxy Technology's EPO-TEK OG142-112 UV Cure Optical Epoxy). To establish the necessary electrical connections from the filter board to the ion trap, a wire bonding tool (2TPT's HB10 Wedge and Ball Bonder) is employed. The atom chip is coated with a top layer of gold, which serves as a protective layer for the wires as well as a mirror for the mMOT laser beams, and is also connected to the system's ground.

\section{\label{sec:RbSetup}Rubidium setup and laser cooling}
In order to achieve the mMOT, it is essential to have laser light and a quadrupole magnetic field. We utilize right-hand circular (RHC) and left-hand circular (LHC) polarized light to drive the $\sigma^{+}$ and $\sigma^{-}$ transitions of the atoms. The optical components of our MOT include two self-constructed external cavity diode lasers (ECDLs) that have a minimum output of 130\,mW under continuous wave conditions (Panasonic LNC728PS01WW). The laser configuration incorporates a reflective grating (GH13-18V: Visible Reflective Holographic Grating, 1800/mm, 12.7\,mm x 12.7\,mm x 6\,mm) placed in Littrow configuration ~\cite{hawthorn2001littrow}. Additionally, this laser design comprises a collimation tube and an aspheric lens (C230TMD-A: $f = 4.51$\,mm, $\text{NA} = 0.55$, Mounted Aspheric Lens, ARC: 350\,-\,700\,nm) to collimate the outgoing laser beam (LT110P-B: $f = $6.24\,mm, $\text{NA} = 0.40$, AR Coated: 650\,-\,1050\,nm) and a decoupling mirror (Tafelmeyer float glass HR/11 E). The laser housing is sealed with an aperture window (WG11050-A: N-BK7 Broadband Precision Window, AR Coated: 350 - 700nm, $t = 5$\,mm). A Peltier element (Quick-Cool QC-71-1.4-8.5M) beneath the laser diode mount stabilizes the temperature of the laser diode. To adjust the laser frequency to the desired values, we vary the voltage applied to the piezo behind the reflective grating, which changes the length of the external cavity. Additionally, the temperature and current of the laser diode can be adjusted.

Prior to initiating the atom trap, the assembled vacuum chamber is opened and a 3D Hall probe (Teslameter FM302 AS-L3DM) is utilized to measure the magnetic fields along the axial and one of the radial directions. The magnetic field gradient is found to be below the expected value due to slight angular deviations in the MOT coil wires resulting from the winding process, as well as a deviation from a perfect circular shape. Additionally, the center of the magnetic field is observed to be a few millimeters away from the geometric center of the chamber.

In the final experiment, the objective is to optimize the loading of the trap with a significant number of atoms in the mirror optical magnetic trap (mMOT). The atoms are laser cooled and confined in a large area mMOT. Once the atoms are captured in the mMOT, the mMOT coils will be transitioned to a bias field while activating the large U-shaped wire beneath the atom chip which generates a quadrupole field with a minimum at infinity. This allows the atoms to be confined in a spatially reduced area, approximately 2\,mm beneath the chip surface. The atoms will then be transferred to a potential created by the small u-shaped wire on the atom chip. At this stage, the atoms are shifted close to the ion trap surface and the atom cloud is compressed to a smaller and steeper mMOT volume so that atoms couple to the atom chip. The final step involves creating an Ioffe-Pritchard trap by activating the z-shaped wire on the atom chip in conjunction with a bias magnetic field. The axial direction of the atom cloud trapped in the magnetic field of the z-shaped wire is not necessarily parallel to the z-axis of the ion trap. In the atom trap design, the length of the z-shaped wire is considered to be 1.4\,mm, so that the atom cloud overlaps with the ion cloud.

As the initial step to activate the atom trap, the magnetic field and rubidium laser system are calibrated. Subsequently, a cloud of rubidium atoms is observed on the EMCCD camera. Under typical experimental conditions (13\,G/cm-axial magnetic field gradient, 2$\pi\times$12\,MHz cooling laser detuning), we acquire a cloud with an approximate radius of $r = ,\sim (1.8\pm0.2)$\,mm that contains roughly (8.7$\pm2)\times$10$^7$ $^{87}$Rb atoms. These values are determined from fluorescence signals.

		\begin{figure}
		\centering
		\includegraphics[scale=0.5]{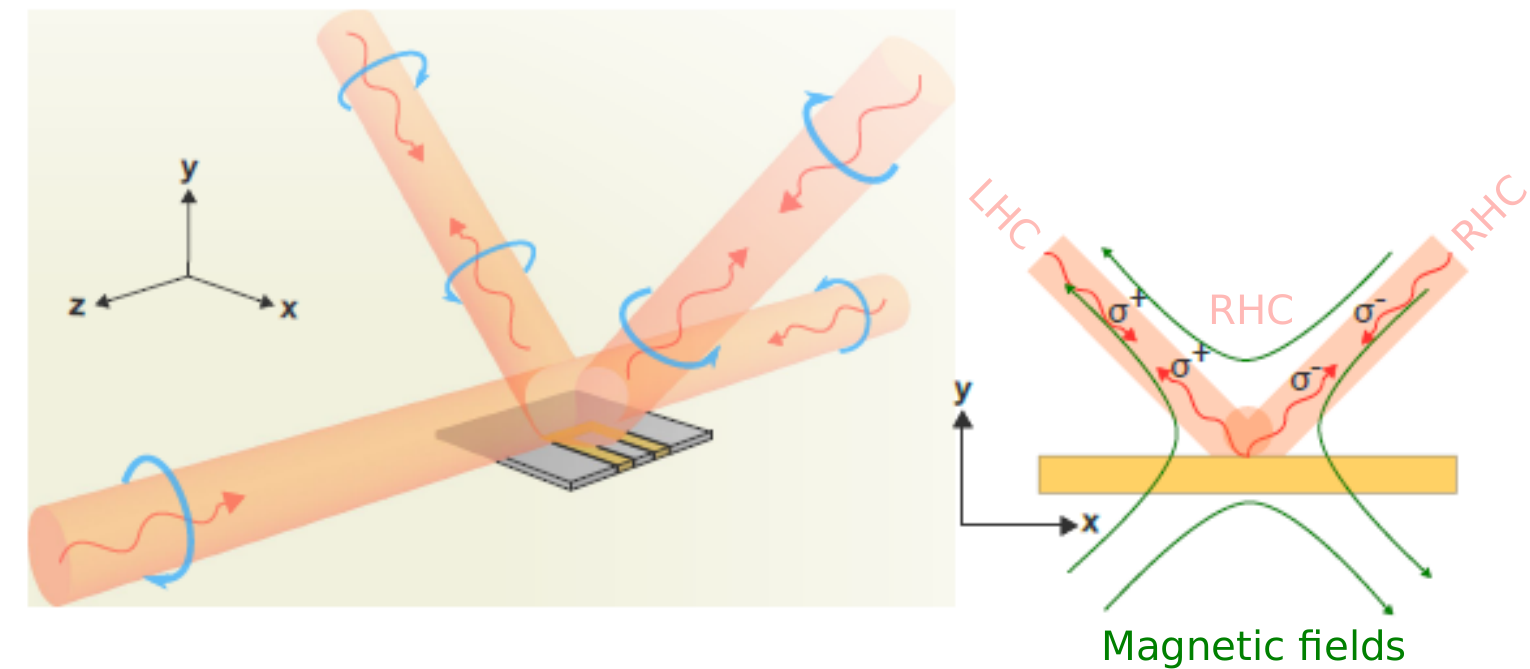}
		\caption{Laser beams (red) with RHC and LHC laser lights to run the $\sigma^{+}$ and $\sigma^{-}$ transitions with a quadrupole magnetic field (green).}
		\label{fig:QPbeams}
		\end{figure}

\subsection{Internally water-cooled MOT coils}

\begin{figure}
	\centering
	\includegraphics[scale=0.6]{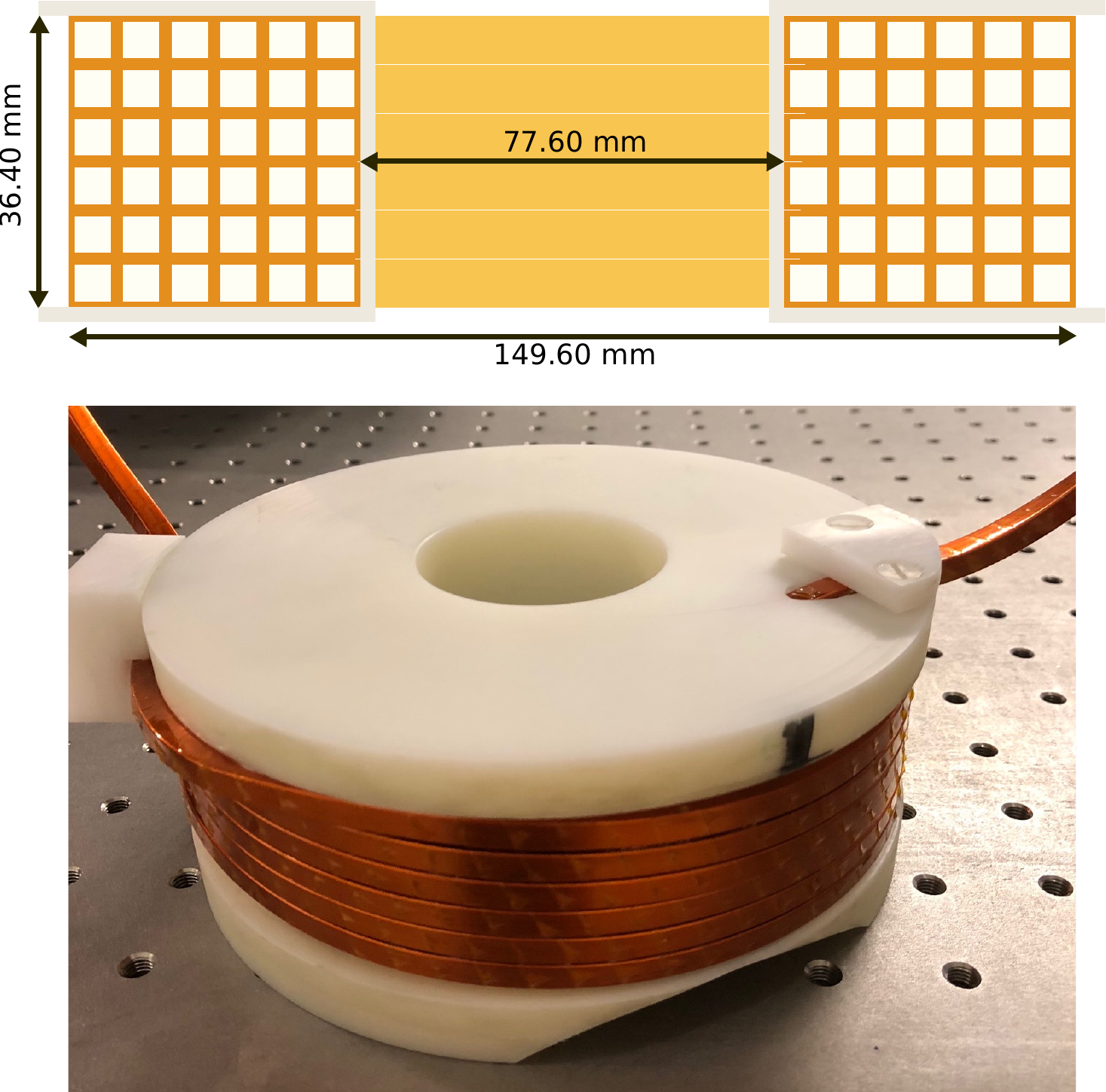}
	\caption{a) Sketch of one of the MOT coils. Each coil comprises 6$\times$6 turns made of a hollow-core copper wire with an external (internal) cross-sectional area of 6$\times$6\,mm$^{2}$ (4$\times$4\,mm$^{2}$); b) Photograph of a MOT coil, the wires are electrically insulated with the use of Kapton tape.}
	\label{fig:MOTcoils}
	\end{figure}

The quadrupol magnetic field needed for the mMOT must be specifically engineered. A field that increases in strength as the distance from its center increases is necessary. In this text, we discuss the creation of a quadrupole magnetic field using two coils in an anti-Helmholtz configuration (Fig. \ref{fig:MOTcoils}). We constructed a set of MOT coils using hollow-core copper wire with a cross-sectional area of 6$\times$6\,mm$^{2}$ (inner cross-sectional area 4$\times$4\,mm$^{2}$). To ensure proper insulation between the wires, they were wrapped twice in Kapton tape. Each coil comprises 36 turns and has an internal diameter of 77.60\,mm and an external diameter 149.60\,mm. During MOT operation, the maximum current of $I=200$\,A provided by a power supply (SM30-100D DELTA elektronika power supply SM 30-200) flows through the wires of the coils. This current is sufficient to generate an axial (radial) magnetic field gradient of 0.059$\times I$[A]\,G/cm (0.029$\times I$[A]\,G/cm) in the vicinity of the trap center, as can be seen in Tab. \ref{Tab:MOT_coils}.

\begin{table}
\begin{center}
\begin{tabular}{l c r}

\textbf{Quantity} & \textbf{Value} & \textbf{Unit}\\
\hline

 Number of winding 	&	36		    &			\\
 Cross-section area	&	20.0	    &	mm$^{2}$\\
 Coil length		&	11.99	    &	m 		\\
 Coil mass			&	2.148	    &	Kg		\\
 Water mass		    &	0.1913	    &	Kg		\\
 Coil resistance	&	0.0100	&	$\Omega$ \\
 Power dissipation	&	419.081	&	W		\\
 Voltage drop		&	2.05432	&	V		\\
 Pressure drop		&	0.5		&	bar 		\\
 Volumetric flow rate	&	2.36196	&	\nicefrac{l}{min}	\\
 Water mass flow rate &	2.35667	&	\nicefrac{kg}{min}	\\
 Fluid velocity		&	2.46038	&	\nicefrac{m}{s} 		\\
 Reynolds number	&	2575.4	&			\\
 Temperature rise	&	2.550	&	$^\circ$C 		\\
\end{tabular}
\caption{Details of the individual MOT coils' production.}
\label{Tab:MOT_coils}
\end{center}
\end{table}

\begin{figure}
	\centering
	\includegraphics[scale=0.9]{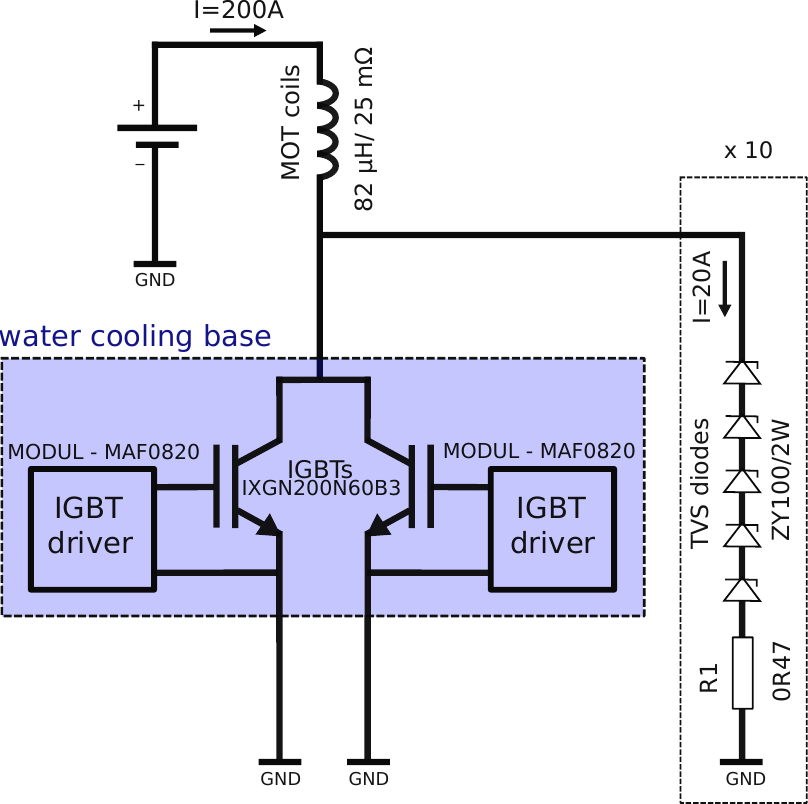}
	\caption{The circuit diagram for rapidly shutting off the coil current is illustrated below. The magnetic trap coils are connected in series, resulting in a total electrical inductance of 82\,$\mu$H and a resistance of 25\,m$\Omega$. When the maximum current of 200\,A is switched off, an electromotive force is generated that triggers the transient-voltage-suppression diodes.}
	\label{fig:switch}
	\end{figure}

To estimate the thermal budget of the coils, we calculated the volumetric flow rate of the cooling water which is an incompressible liquid ~\cite{cornish1928flow}. The maximum pressure drop in our chiller is 4.5\,bar (Van der Heijden MINORE 0-RB400). With a pressure loss of about 0.5\,bar, we estimate a temperature increase of about 2-4$^\circ$C, which is in close agreement with the measured temperature increase when the MOT coils were running continuously at $I=200$\,A. At this current, the energy dissipated in each coil is approximately 900\,Watt and the voltage loss 4.15\,V. The voltage loss increases to 7\,V when we engage the MOT switch (Fig.~\ref{fig:switch}).



\subsection{Rapid high-current MOT switch}
 Fast magnetic field switching is essential for maintaining the stability of the MOT, enhancing the laser cooling efficiency, and for performing various manipulation of trapped atoms. However, doing so caused eddy currents to form in the electrical conductive parts, leading to a slow decay of the magnetic field. To solve this issue, we implemented a Polyoxymethylene holder for the coils. To achieve fast switching, we developed a current driver using high-speed insulated gate bipolar transistors (IGBTs). We utilized a series connection of 10$\times$5 transient-voltage-suppressor diodes (TVS diodes) followed by a resistor. Each diode has a breakdown voltage of 100\,V, allowing the magnetic energy to dissipate to ground as soon as the reverse voltage reaches 500\,V. With this setup, we succeeded in achieving a switching-off time less than 100\,$\mu$s for 200\,A. We observed a linear relationship between the applied current in the magnetic trap coils and the switching-off time, with a rate of 0.45\,$\mu$s per 1\,A.

\subsection{Operation of the mMOT}
The mMOT is based on the reflection of trapping light beams off a plane mirror. In our setup, the gold surface of the planar chip trap is reflecting the beams. Initially we tested the mMOT in a test setup with just a gold mirror (Thorlabs PF20-03-M03). We aligned all the laser beams with the zero of the quadrupole magnetic field and trapped a mMOT directly from the background vapor of the atoms. Then, we resumed to the beam reflection by the final chip carrier including the atom-ion chip. By decreasing the diameter of the MOT beams from 10\,mm to 5\,mm, we achieved trapping of a very small sample of $\sim 3\times 10^{7}$ atoms of $^{87}$Rb, 2\,mm below the ion chip area, thus demonstrating a loading method for the magnetic trap which is formed from the magnetic field of the z-shaped wire.

				\begin{figure}[h]
				\centering
				\includegraphics[scale=0.8]{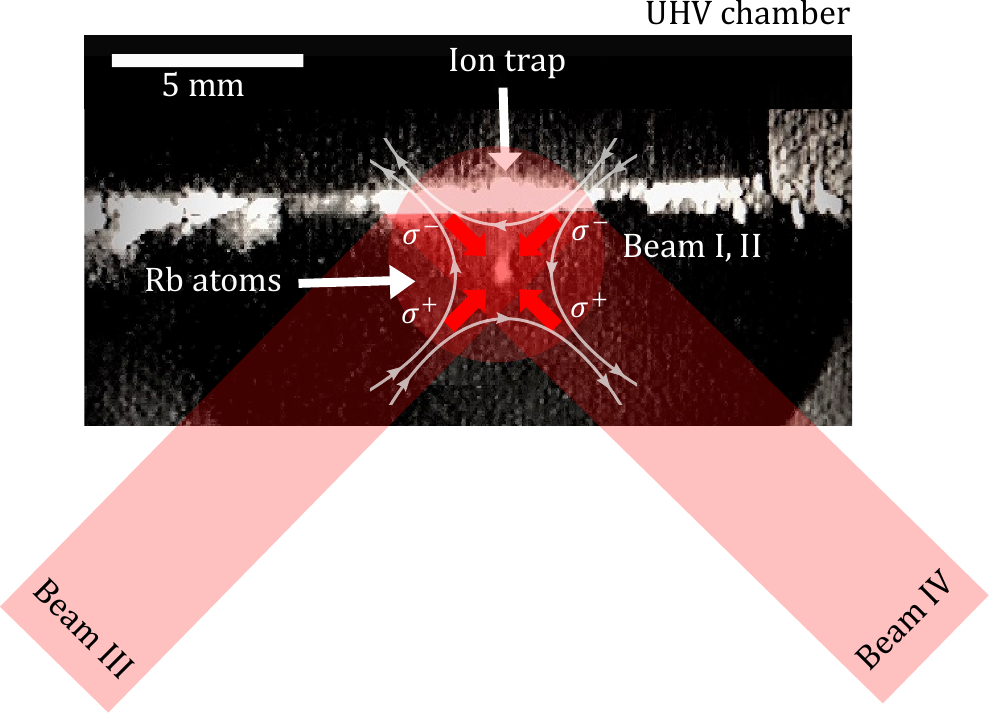}
				\caption{First image of the $^{87}$Rb MOT under the ion chip. Beams number I, II and III are all RHC polarized while beam IV is LHC to run the $\sigma ^{\pm}$ transitions. The size of the MOT is smaller compared to the MOT trapped under the gold mirror. This is due to the microfabricated structures on the ion chip surface and a 100\,$\mu$ gap in the middle of the chip. Additionally, the ion chip distorts the shape of the reflected beams coming off of the chip surface. The ion trap, with a width of $w=5$\,mm, is also visible due to light reflections.}
				\label{fig:mMOT}
				\end{figure}

\section{Conclusion and Future Directions}
The integrated hybrid atom-ion setup described in this paper offers several advantages over other hybrid atom-ion setups currently available. One major advantage is the compactness and accessibility provided by the chip-based design, which allows for easy integration with other experimental apparatus and simplifies the process of manipulating and studying both atoms and ions. Additionally, the precise mutual positioning of atoms and ions within the device enables more accurate measurements and control over the interactions between these particles. Furthermore, the ability to cool one component using the other component can lead to improved precision and control in the experiments. The compactness and integration of the setup could also be the key to make it a more cost-effective option compared to other existing hybrid setups. The setup could be easily scaled up to suit different experimental needs. The integration of both atom and ion trapping on a single chip enables new possibilities for precision measurements and quantum computing applications. The ability to trap and manipulate both atoms and ions in a single chip is a significant advancement in the field, and we look forward to the many exciting developments that will result from this technology.

\section{Acknowledgments}
The authors would like to acknowledge the support and contributions of all individuals involved in this research. We are grateful to Rene Gerritsma and Jannis Joger for their invaluable assistance and expertise. Additionally, we would like to thank xxx for their financial support of this project.

\bibliography{apssamp}

\end{document}